\documentstyle[preprint,pra,aps,epsf]{revtex}

\begin{document}
\draft
\title{Radiative association and inverse predissociation\\
of oxygen atoms}
\author{J. F. Babb and A. Dalgarno}
\address{
Harvard-Smithsonian Center for Astrophysics, 60 Garden Street,
Cambridge, MA 02138}
\date{\today}
\maketitle
\begin{abstract}
The formation of $\mbox{O}_2$ by radiative association and by inverse
predissociation of ground state oxygen atoms is studied using
quantum-mechanical methods.  Cross sections, emission spectra, and
rate coefficients are presented and compared with prior experimental
and theoretical results.  At temperatures below 1000~K radiative
association occurs by approach along the $1\,{}^3\Pi_u$ state of
$\mbox{O}_2$ and above 1000~K inverse predissociation through the
$\mbox{B}\,{}^3\Sigma_u^-$ state is the dominant mechanism. This
conclusion is supported by a quantitative comparison between the
calculations and data obtained from hot oxygen plasma spectroscopy.
\end{abstract}

\narrowtext
\section{Introduction}
\begin{figure}
\epsfxsize=1.0\textwidth \epsfbox[0 0 600 612]{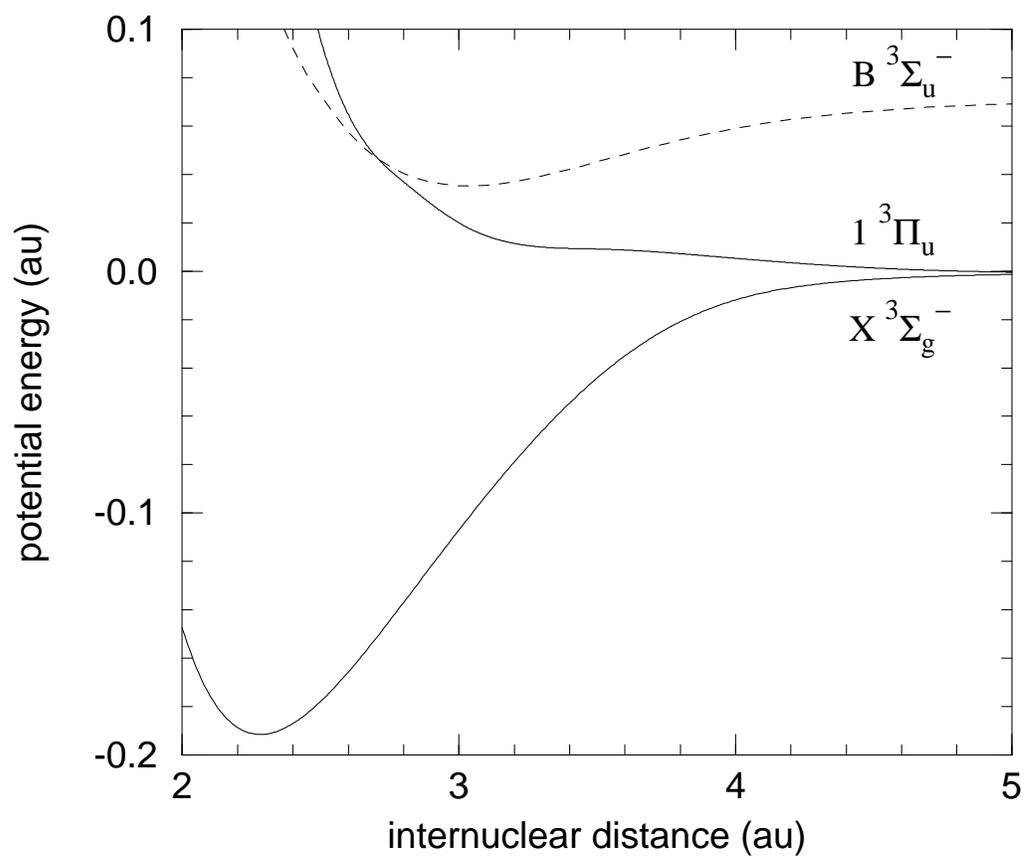}
\caption{\label{potentials}Adopted $1\,{}^3\Pi_u$, $\mbox{B}\,{}^3\Sigma_u^-$,
and $\mbox{X}\,{}^3\Sigma_g^-$ potentials.}
\end{figure}

Of the 81 molecular electronic states that two
$\mbox{O}({}^3\mbox{P})$ atoms can form, the repulsive $1\,{}^3\Pi_u$
state can make an allowed electric dipole transition to the
$\mbox{X}\,{}^3\Sigma_g^-$ ground state.  Thus, the formation of an
oxygen molecule through the {\em direct radiative association\/}
reaction
\begin{equation}
\label{direct-reaction}
\mbox{O}({}^3\mbox{P}) + \mbox{O}({}^3\mbox{P})
 \rightarrow \mbox{O}_2(1\,{}^3\Pi_u)
\rightarrow \mbox{O}_2(\mbox{X}\,{}^3\Sigma_g^-) + h\nu
\end{equation}
can occur and it might be an important molecular oxygen creation
process at low temperatures.  In Fig.~\ref{potentials} the potential
energy curves for the $1\,{}^3\Pi_u$ and $\mbox{X}\,{}^3\Sigma_g^-$
states of $\mbox{O}_2$ are depicted.  Wraight~\cite{Wra77} had
suggested that the process~(\ref{direct-reaction}) might be
responsible for the emission in the near-infrared of the nightglow of
the planet.

At higher temperatures another process---{\em inverse
predissociation\/}, an indirect mechanism of association---can occur
with the excited $\mbox{B}\,{}^3\Sigma_u^-$ state serving as an
intermediate step in the formation of an oxygen molecule
\begin{equation}
\label{indirect-reaction-general}
\mbox{O}({}^3\mbox{P}) + \mbox{O}({}^3\mbox{P})
 \rightarrow \mbox{O}_2^* \rightarrow \mbox{O}_2(\mbox{B}\,{}^3\Sigma_u^-)
 \rightarrow \mbox{O}_2(\mbox{X}\,{}^3\Sigma_g^-) + h\nu ,
\end{equation}
where $\mbox{O}_2^*$ denotes any one of the four states $1\,{}^1
\Pi_u$, $1\,{}^3\Pi_u$, $1\,{}^5 \Pi_u$, or $2\,{}^3 \Sigma_u^+$ that
dissociate to $\mbox{O}({}^3\mbox{P})+\mbox{O}({}^3\mbox{P})$ and
cross the $\mbox{B}\,{}^3\Sigma_u^-$ state.  The predissociative
coupling is due to spin-orbit (any of the four states) or
electronic-rotational interaction ($1\,{}^3\Pi_u$
state)~\cite{JulKra75}.  Reaction~(\ref{indirect-reaction-general})
may be characterized by an activation energy that is the energy with
respect to the $\mbox{O}({}^3\mbox{P})+\mbox{O}({}^3\mbox{P})$
asymptote at which the particular state $\mbox{O}_2^*$ crosses the
$\mbox{B}\,{}^3\Sigma_u^-$ state.  The potential energy curve of the
$\mbox{B}\,{}^3\Sigma_u^-$ state, which dissociates to an
$\mbox{O}({}^3\mbox{P})$ atom and an $\mbox{O}({}^1\mbox{D})$ atom is
shown in Fig.~\ref{potentials}.  Because the strength of the
electronic-rotational coupling increases with the nuclear rotational
angular momentum, while the strength of the spin-orbit coupling is
nearly independent of it~\cite{LefBriFie86}, the $1\,{}^3\Pi_u$ state
is expected to be the most important channel for inverse
predissociation.  In a shock tube study of the emission from
recombining oxygen atoms in a non-equilibrium shocked ozone-argon gas
mixture at temperatures between 2500 and 3800~K Myers and
Bartle~\cite{MyeBar68} obtained absolute intensity measurements of the
spectrum, and attributed the emission to the formation of oxygen
molecules from ground state oxygen atoms through the inverse
predissociation process
\begin{equation}
\label{indirect-reaction}
\mbox{O}({}^3\mbox{P}) + \mbox{O}({}^3\mbox{P})
 \rightarrow \mbox{O}_2(1\,{}^3\Pi_u)
 \rightarrow \mbox{O}_2(\mbox{B}\,{}^3\Sigma_u^-)
 \rightarrow \mbox{O}_2(\mbox{X}\,{}^3\Sigma_g^-) + h\nu .
\end{equation}
The emission has been observed in several later experiments.  It was
seen in a silent electric discharge in $\mbox{O}_2$ by Weisbeck and
V\"olkner~\cite{WeiVol71} and Sharma and Wray~\cite{ShaWra71} measured
emission intensities at 230 and 325~nm for temperatures in the range
2800--5300~K in shocked oxygen and oxygen-noble gas mixtures.  Sharma
and Wray~\cite{ShaWra71} confirmed that the excited levels of
$\mbox{O}_2$ were populated by inverse predissociation but their
emission spectra appear to be modified by vibrational redistribution.
Similar conclusions were reached by Wray and Fried~\cite{WraFri71} who
measured the emission intensities in an arc jet in oxygen and and
oxygen-like gas mixtures.  Hoffmann and Neiger~\cite{HofNei78,Hof79}
measured the absolute intensity of the emission from an equilibrium
oxygen plasma at 8800, 9140, and 9420~K, finding a peak at about
260~nm which they attributed to inverse
predissociation~(\ref{indirect-reaction}).  We return to a discussion
of prior work in Sec.~\ref{sec:results}.

In this paper, we study quantum-mechanically the direct radiative
association process~(\ref{direct-reaction}) and the inverse
predissociation process~(\ref{indirect-reaction}).  Cross sections,
emission spectra, and rate coefficients are determined for
temperatures from 15 to $15\,000$~K.  Improvements in the knowledge of
the $1\,{}^3\Pi_u$ and $\mbox{B}\,{}^3\Sigma_u^-$ state potentials
make it possible to get qualitative agreement with many experiments
for the emission spectra and quantitative agreement for rate
coefficients measured by Myers and Bartle~\cite{MyeBar68}.

\section{Molecular states}
\label{molecular-properties}

In Ref.~\cite{BabSunJam93} we constructed two $1\,{}^3\Pi_u$ potential
energy curves that were identical except in the range of internuclear
distances $2.4<R<3.1$ (we shall use atomic units in this section and
Sec.~\ref{direct-section}).  The first curve was designed to be
consistent with the level shifts arising from predissociation of the
$\mbox{B}\,{}^3\Sigma_u^-$ state, as analyzed in
Ref.~\cite{ChiCheFin92} where an internuclear distance for the
$\mbox{B}\,{}^3\Sigma_u^-$- $1\,{}^3\Pi_u$ crossing $R_x=2.700$ was
obtained, and which used {\em ab initio\/} energy
calculations~\cite{ParBauLan91} for $3.1<R<10$.  The second curve we
constructed simply used the {\em ab initio\/}
calculations~\cite{ParBauLan91} for $2.5<R<10$.  A recent
analysis~\cite{LewGibDoo94} of the photoabsorption cross section
measurements~\cite{YosEsmChe92}, an analysis presumably more accurate
than that of Ref.~\cite{ChiCheFin92}, yielded a different crossing
point of $R_x=2.748$, which is consistent with the crossing point of
the cubic spline fit~\cite{LewGibDoo94} to the calculations
of~\cite{ParBauLan91} at $R=2.5$ and 2.8.  We have therefore used the
second curve we constructed in Ref.~\cite{BabSunJam93} here.  As we
have discussed~\cite{BabSunJam93} the potential curve was supplemented
for $R<2.4$ by the adjusted {\em ab initio\/} energies
of~\cite{AllGubDal82} and was constructed to have the proper
asymptotic long-range behavior.

The adopted transition dipole moment function $D(R)$, defined
in~\cite{GubDal79}, for the transition from the $1\,{}^3\Pi_u$ to the
$\mbox{X}\,{}^3\Sigma_g^-$ state was the semi-empirical determination
of ~\cite{WanMccBla87} for $1.5118 \leq R \leq 2.5$ and the {\em ab
initio\/} calculation of~\cite{AllGubDal82} for $2.6\leq R\leq 5$.
The data were smoothly connected using cubic splines. For $R \leq
1.5118$ the function $D(R)=0.1794 - 0.01945R$ was used and for $R\geq
5$ the function $D(R)=1.2625/R^3$.  The $\mbox{X}\,{}^3\Sigma_g^-$
state potential energy curve was constructed using RKR data for
$1.79480 < R < 4.0840$ and fitted to the long-range form $ -17.57/R^6
$ for $R>5$ as described in Ref.~\cite{Fri90}, but with a short-range
form
\begin{equation}
V(R) =  5757.0053\exp(-5.7925 R) .
\end{equation}
The adopted $\mbox{X}\,{}^3\Sigma_g^-$ and $1\,{}^3\Pi_u$ potential
curves, which correlate with the separated atoms
$\mbox{O}({}^3\mbox{P}) + \mbox{O}({}^3\mbox{P})$, are shown in
Fig.~\ref{potentials}.  The $\mbox{B}\,{}^3\Sigma_u^-$ potential curve
of~\cite{Fri90} and the semi-empirical
$\mbox{B}\,{}^3\Sigma_u^-$--$\mbox{X}\,{}^3\Sigma_g^-$ transition
dipole moment of~\cite{AllGubDal86} as described in~\cite{Fri90} were
utilized.  In Fig.~\ref{potentials} the adopted
$\mbox{B}\,{}^3\Sigma_u^-$ potential is shown.  It correlates to the
separated atom limit $\mbox{O}({}^3\mbox{P}) + \mbox{O}({}^1\mbox{D}$,
which lies 0.07229~au (1.967~eV) above $\mbox{O}({}^3\mbox{P})
+\mbox{O}({}^3\mbox{P})$.

\section{Direct radiative association}
\label{direct-section}
A fully quantum-mechanical theory can be formulated and used to
calculate the cross sections at various energies and rate coefficients
at various temperatures for direct radiative
association~(\ref{direct-reaction}).  Let $E$ be the energy of
relative motion of the atoms approaching in the $1\,{}^3\Pi_u$
electronic state.  The cross section for a transition to a bound
vibration-rotation level $E_{{v''}{N''}}$ of the
$\mbox{X}\,{}^3\Sigma_g^-$ molecular ground state is given in atomic
units by~\cite{Bat51,Pal67}
\begin{equation}
\label{direct-partial-cross}
\sigma_{{v''}{N''}N} (E) = \frac{g_1}{g}
 \frac{4\pi^2\alpha^3}{3\mu E}
  (E + E_{{v''}{N''}})^3
  S_{J'',J} |M_{{N''},N}|^2
\end{equation}
where $\mu=14583.10$ is the reduced mass for ${}^{16}\mbox{O}_2$, $g_1
=3$ is the statistical weight of the $1\,{}^3\Pi_u$ state (the nuclear
spin is zero so only one of each lambda-doubling level is populated)
and $g=81$ is the statistical weight of the
$\mbox{O}({}^3\mbox{P})$--$\mbox{O}({}^3\mbox{P})$ pair, $\alpha$ is
the fine structure constant, $S_{J'',J}$ is the H\"onl-London factor,
and
\begin{equation}
M_{{N''},N}
   = \int_0^\infty dR \,\psi_{{v''}{N''}} (R)  \, D(R) \, \chi_{N}(R) ,
\end{equation}
with $\psi_{{v''}{N''}} (R)$ the normalized bound state wave function
and $\chi_{N}(R)$ the energy-normalized wave function for the partial
wave with angular momentum $N$ of the continuum $1\,{}^3\Pi_u$ state.
The wave functions $\psi_{{v''}{N''}}$ and $\chi_{N}$ were calculated
by integrating the Schr\"odinger equation using the Numerov method and
$\chi_{N}$ was matched to the asymptotic form
\begin{equation}
\label{asymptotic}
\chi_{N}(R) \sim (2\mu/\pi k)^{1/2} \sin( kR + \delta_N
             -\case{1}/{2}N\pi ) ,\qquad R\sim\infty ,
\end{equation}
with $k=(2\mu E)^{1/2}$ and $\delta_N$ the phase shift.  Due to the
presence of identical nuclei in the diatom and the absence of nuclear
spin, the $\mbox{X}\,{}^3\Sigma_g^-$ state is allowed only odd values
${N''}$.  Let $J=|{\bf N}+{\bf S}|$ be the magnitude of the total
angular momentum for the $1\,{}^3\Pi_u$ state, with ${\bf S}$ the
total spin angular momentum and $|{\bf S}|=1$. Assuming that Hund's
case~(b) coupling applies, we have for $N\gg 1$, $J\sim N$ (in an
unrelated study, case~(b) was found to be a satisfactory approximation
to intermediate coupling~\cite{Fri90}) and then, summing over the
rotational branches~\cite{NolJen36} for the allowed fine structure
transitions for a given value of $N$ we obtain the line strength
values given in Table~\ref{line-strength-table}, which possess the sum
rule property~\cite{Sch78,WhiSchTat80}
\begin{equation}
\sum_{N''} S_{N'',N} = (2S+1)(2N +1) .
\end{equation}
In carrying out the computations $N''$ was required to be odd, but $N$
could be odd or even.

\begin{figure}
\epsfxsize=1.0\textwidth \epsfbox[0 0 600 612]{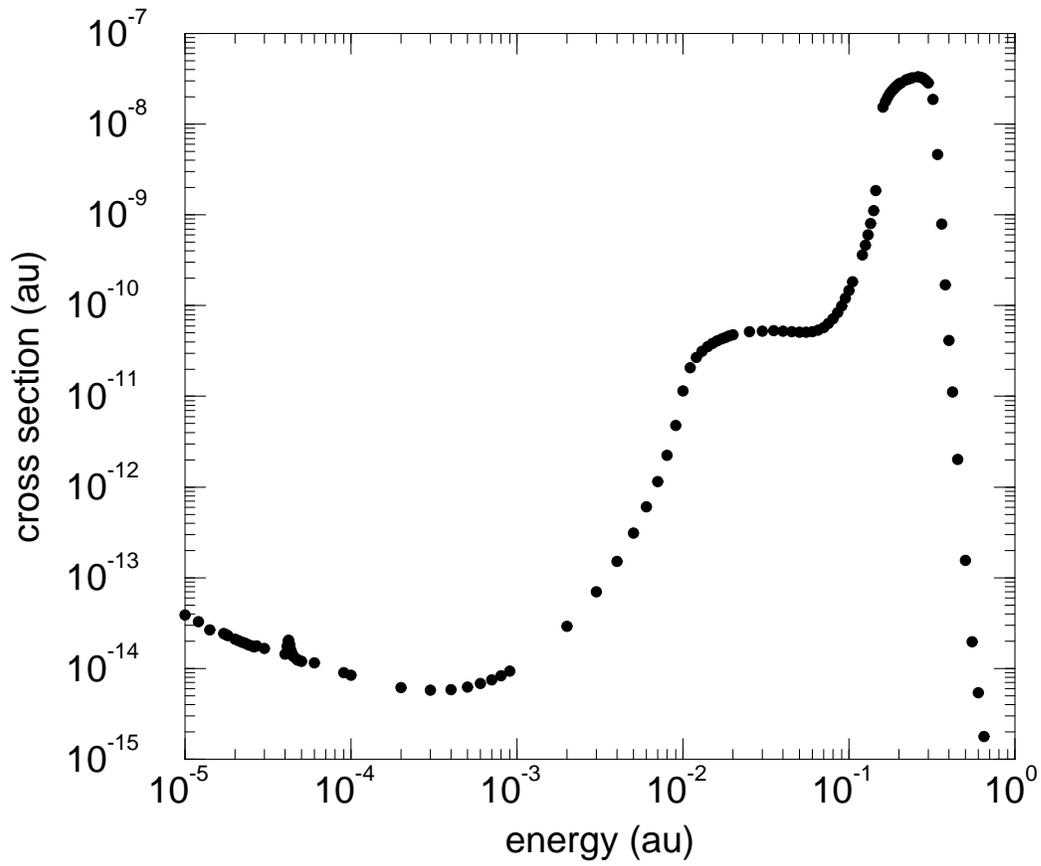}
\caption{\label{direct-cross}Cross section
$\sigma(E)$~(\protect\ref{total-cross})
in au $(a_0^2)$ for the direct radiative association
process~(\protect\ref{direct-reaction}) as a function of energy~$E$
in au $(e^2/a_0)$.
}
\end{figure}

The total cross section, summed over allowed transitions between
partial waves $N$ and final states ${v''}{N''}$,
\begin{equation}
\label{total-cross}
\sigma(E) \equiv
      \frac{1}{(2S+1)} \sum_{{v''}{N''}N} \sigma_{{v''}{N''}N} (E)
\end{equation}
is presented in Fig.~\ref{direct-cross} at various energies.  Some
structure, due to shape resonances, is apparent in the cross section
at low energies.  The cross section drops off rapidly for $E \agt 0.4$
due to the loss of Franck-Condon overlap at small $R$.  In computing
the total cross sections values of $N$ up to 15 were sufficient for
$E\alt 2.0\times 10^{-5}$, up to 30 for $E \alt 3.0\times 10^{-5}$, up
to 100 for $E \alt 0.015$, and for $E \agt 0.015$ values of $N$ up to
188 were used.  Since the $\mbox{X}\,{}^3\Sigma_g^-$ state has 45
vibrational levels each total cross section at a given $E$ was a sum
of many partial cross sections.  Nevertheless, the computations could
be carried out readily.

The cross sections obtained at various values of $E$ were used to
obtain the rate coefficient at various temperatures by averaging over
a Maxwellian velocity distribution,
\begin{equation}
\label{Max-Boltz}
\alpha_{\rm D}(T) = \frac{2}{kT} \left( \frac{2}{\pi\mu kT} \right)^{1/2}
          \int_0^\infty E \sigma (E) e^{-E/kT} \, dE .
\end{equation}
The values are given in column~2 of Table~\ref{rate-table}.  The rate
coefficient is approximated to within 5\% for $T<400$~K by the
expression
\begin{equation}
  \alpha_{\rm D}(T) \approx  9\times10^{-27}T^{-0.23}\exp(T/70) ,\quad
      T<400~\mbox{K}.
\end{equation}

Wraight~\cite{Wra77} estimated semiclassically the rate coefficient
for the direct process~(\ref{direct-reaction}) and found
\begin{equation}
\label{Wraight-rate}
\alpha_{\rm W}(T)
\sim 6 \times 10^{-24} \, T^{1.58} \;  \mbox{cm}^3\mbox{s}^{-1} ,
\end{equation}
for $200~\mbox{K}<T<2000~\mbox{K}$. He advised caution in applying
(\ref{Wraight-rate}) to high temperatures.
Expression~(\ref{Wraight-rate}) was adopted for astrophysical purposes
in the tabulation of Prasad and Huntress~\cite{PraHun80} and in the
UMIST data base~\cite{MilRawBen91}.  The present computations show
that Eq.~(\ref{Wraight-rate}) drastically overestimates the radiative
association rate coefficient at low temperatures.

\section{Inverse predissociation}
For inverse predissociation, (\ref{indirect-reaction}), the rate
coefficient at temperature $T$ is given by the Breit-Wigner
expression~\cite{BaiBar72,GolThr73,DuDal90}
\begin{equation}
\label{IP-def}
\alpha(T) = \frac{g_{\rm B}}{g} \left( \frac{2\pi}{\mu kT} \right)^{3/2}
\; \sum_{{v'}} \sum_{{N'}} \;
(2{N'}+1)
\frac{\Gamma_{\rm r}({v'},{N'})
         \Gamma_{\rm d}({v'},{N'})}
      {\Gamma_{\rm r}({v'},{N'})+
         \Gamma_{\rm d}({v'},{N'})}
\exp ( -E_{{v'}{N'}}/kT) ,
\end{equation}
where $g_{\rm B}=3$ is the statistical weight of the
$\mbox{B}\,{}^3\Sigma_u^-$ state, $g=81$ is the statistical weight of
the $\mbox{O}({}^3\mbox{P})$--$\mbox{O}({}^3\mbox{P})$ pair,
$E_{{v'}{N'}}$ is the energy of the level ${v'}{N'}$ of the
$\mbox{B}\,{}^3\Sigma_u^-$ state measured with respect to the
$\mbox{O}({}^3\mbox{P})$--$\mbox{O}({}^3\mbox{P})$ asymptote,
$\Gamma_{\rm r}({v'},{N'})$ and $\Gamma_{\rm d}({v'},{N'})$ are,
respectively, the radiative and predissociative widths of level
${v'}{N'}$, with
\begin{equation}
\label{lifetime}
\Gamma_{\rm r}({v'},{N'})=\hbar A({v'},{N'})
              \equiv \hbar\sum_{{v''}{N''}} A({v'}{N'}{v''}{N''}),
\end{equation}
and $A({v'}{N'}{v''}{N''})$ is the Einstein $A$-coefficient.  The sum
in Eq.~(\ref{IP-def}) is taken over levels ${v'}{N'}$ with energy
greater than the energy of the crossing point.
Analysis~\cite{LewGibDoo94} of the predissociation in the
$\mbox{B}\,{}^3\Sigma_u^-$ state indicates that the $1\,{}^3\Pi_u$
state crosses at an energy of $0.0435$ relative to the
$\mbox{O}({}^3\mbox{P})$--$\mbox{O}({}^3\mbox{P})$
asymptote~\cite{ChiCheFin92} or about $13\,000$~K.  For the
$\mbox{B}\,{}^3\Sigma_u^-$ state, typically $\Gamma_{\rm
d}({v'},{N'})\gg\Gamma_{\rm r}({v'},{N'})$~\cite{CosParCop93}, so that
\begin{equation}
\label{invpre-rate-def}
\alpha(T) \approx \frac{3}{81} \left( \frac{2\pi}{\mu kT} \right)^{3/2}
                                        \; \sum_{{v'}} \sum_{{N'}} \;
            (2{N'}+1) \Gamma_{\rm r}({v'},{N'}) \exp ( -E_{{v'}{N'}}/kT) .
\end{equation}
In evaluating~(\ref{invpre-rate-def}) we made the approximation
\begin{equation}
\label{no-rot}
A({v'}{N'}{v''}{N''})
 \approx  A({v'} 0{v''} 0) [(E_{v'N'}-E_{v''N''})/(E_{v'0}-E_{v''1})]^3,
\end{equation}
with
${N'}\approx{N''}$ leading to the formula
\begin{eqnarray}
\label{invpre-rate}
\alpha(T) &\approx& \frac{3}{81} \left( \frac{2\pi}{\mu kT} \right)^{3/2}
           \; \sum_{{v'}}\sum_{{v''}}\hbar A({v'} 0{v''} 0) \nonumber\\
 & &   \times   \sum_{{N'}}
            (2{N'}+1)
     [(E_{{v''}{N'}}-E_{{v''}{N'}})/(E_{v'0}-E_{v''1})]^3
    \exp ( -E_{{v'}{N'}}/kT) .
\end{eqnarray}

The intensity of emission from a transition from level $({v'},{N'})$
into level $({v''},{N''})$ is then given by the expression
\begin{eqnarray}
\label{int-def}
I_\nu & \approx & h \nu_{{v'}{N'}{v''}{N''}} \case{3}{81}
( 2\pi/\mu kT)^{3/2} (2{N'}+1) \nonumber \\
& &
 \times\exp (-E_{{v'}{N'}}/kT)\hbar
  A({v'} 0{v''} 0)
  [(E_{{v''}{N'}}-E_{{v''}{N'}})/(E_{v'0}-E_{v''1})]^3 .
\end{eqnarray}
With $E({v'}{N'}{v''}{N''})\equiv h \nu_{{v'}{N'}{v''}{N''}}$ and
$E_{{v'}{N'}}$ both in $\mbox{cm}^{-1}$, $T$ in K, and $A({v'}0{v''}
0)$ in $\mbox{s}^{-1}$, we can express $I_\nu$ in W-$\mbox{cm}^3$ as
\begin{equation}
\label{int-watts}
I_\nu = \case{3}{81} (4.7\times 10^{-45})E({v'}{N'}{v''}{N''})
        (1/T^{3/2}) (2{N'}+1) A({v'} 0{v''} 0) \exp (-E_{{v'}{N'}}/0.695T) .
\end{equation}

\section{RESULTS AND COMPARISON WITH EXPERIMENT}\label{sec:results}
\begin{figure}
\epsfxsize=1.\textwidth \epsfbox[0 0 600 612]{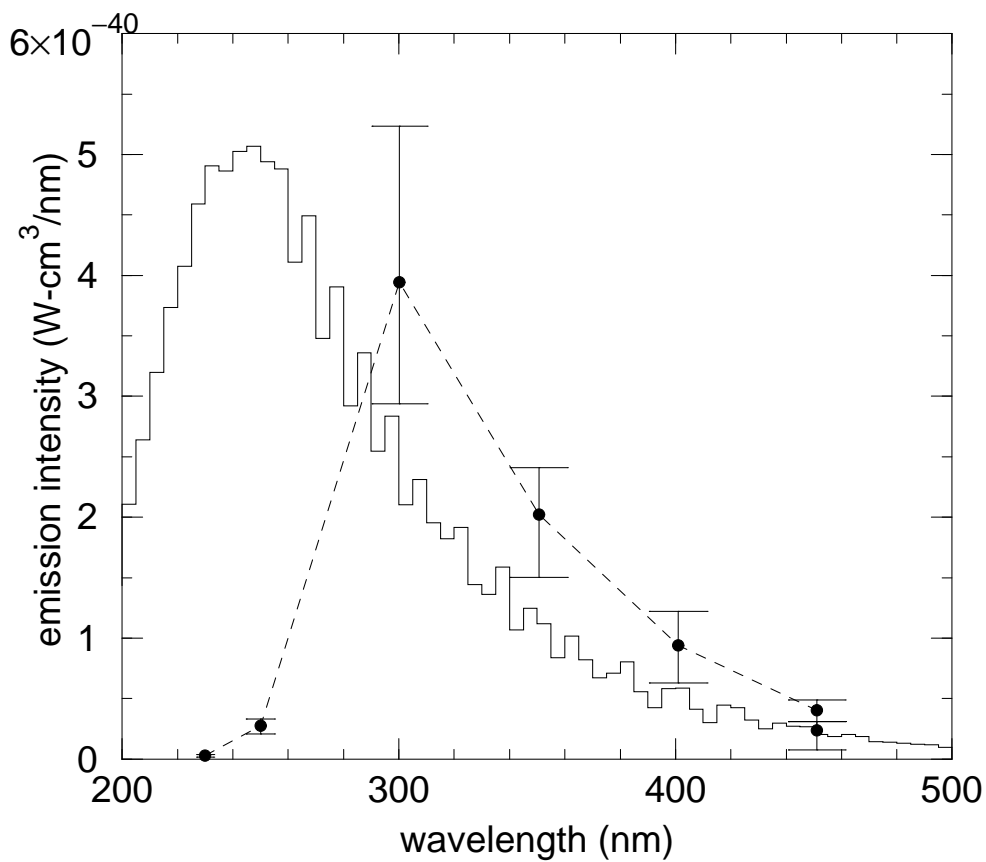}
\caption{\label{IP-spectrum}Absolute emission intensity
at various wavelengths calculated for inverse predissociation at
3000~K (solid line), compared to experimental measurements of Myers
and Bartle at 3030~K (points with error bars connected by the dotted
line).  The calculated spectrum has been averaged over bins of width
5~nm. The experimental data are average values over a spectral
interval the width of which is represented by the horizontal span of
the error bar.}
\end{figure}

Using a program of Allison~\cite{AllDalPas71,AllGubDal86} and the
potentials and transition dipole moments of
Sec.~\ref{molecular-properties} the radiative widths were determined.
They are in good agreement with other
computations~\cite{AllDalPas71,LauKru92}.  In Table~\ref{A-table} we
present the values of the emission transition probabilities
$A({v'},0)$ of the $\mbox{B}\,{}^3\Sigma_u^-$ state vibrational
levels.  In Fig.~\ref{IP-spectrum} the calculated emission spectra
arising in inverse predissociation at 3000~K are shown in
W-$\mbox{cm}^3/\mbox{nm}$ along with the measurements of Myers and
Bartle~\cite{MyeBar68} at 3030~K; the experimental points and error
bars were interpolated from absolute emission-intensity spectral
distribution curves.  Our calculations support the suggestion that the
measurements are in error for wavelengths $\lambda\alt 300$~nm, a
possibility noted by Myers and Bartle~\cite{MyeBar68} and Sharma and
Wray~\cite{ShaWra71}.  The measurements of Sharma and
Wray~\cite{ShaWra71} and Wray and Fried~\cite{WraFri71} appear to be
affected by vibrational redistribution and their spectral data can be
better interpreted in terms of an equilibrium model of the vibrational
level populations of the $\mbox{B}\,{}^3\Sigma_u^-$
state~\cite{All66}.

In the measurements of Hoffmann and Neiger near 9000~K a strong broad
continuum feature was observed at around 260~nm.  Our calculated
emission spectra from inverse predissociation are consistent with the
experimental observations.  Measurements at higher wavelength
resolution should reveal structure arising from the Schumann-Runge
bands populated by inverse predissociation.  We find an absolute
emission intensity of about $5\times10^{-39}$~W-$\mbox{cm}^3$/nm at
250~nm and 9400~K.  From the measured absorption coefficient of
$K=2\times10^{-5}\mbox{cm}^{-1}$~\cite{HofNei78}, using Wien's
radiation law~\cite{Hof79,Pen59}, the length 10~cm of the experimental arc
column, and estimating the number density~\cite{Hof79} of O atoms of
about $10^{18}\,\mbox{cm}^{-3}$ from Fig.~5 of Ref.~\cite{Hof79}, we
obtain an experimental value of $2.5\times10^{-39}$~W-$\mbox{cm}^3$/nm
at 250~nm and 9400~K.  The agreement to within a factor of two is
remarkably close.

\begin{figure}
\epsfxsize=1.0\textwidth \epsfbox[0 0 600 612]{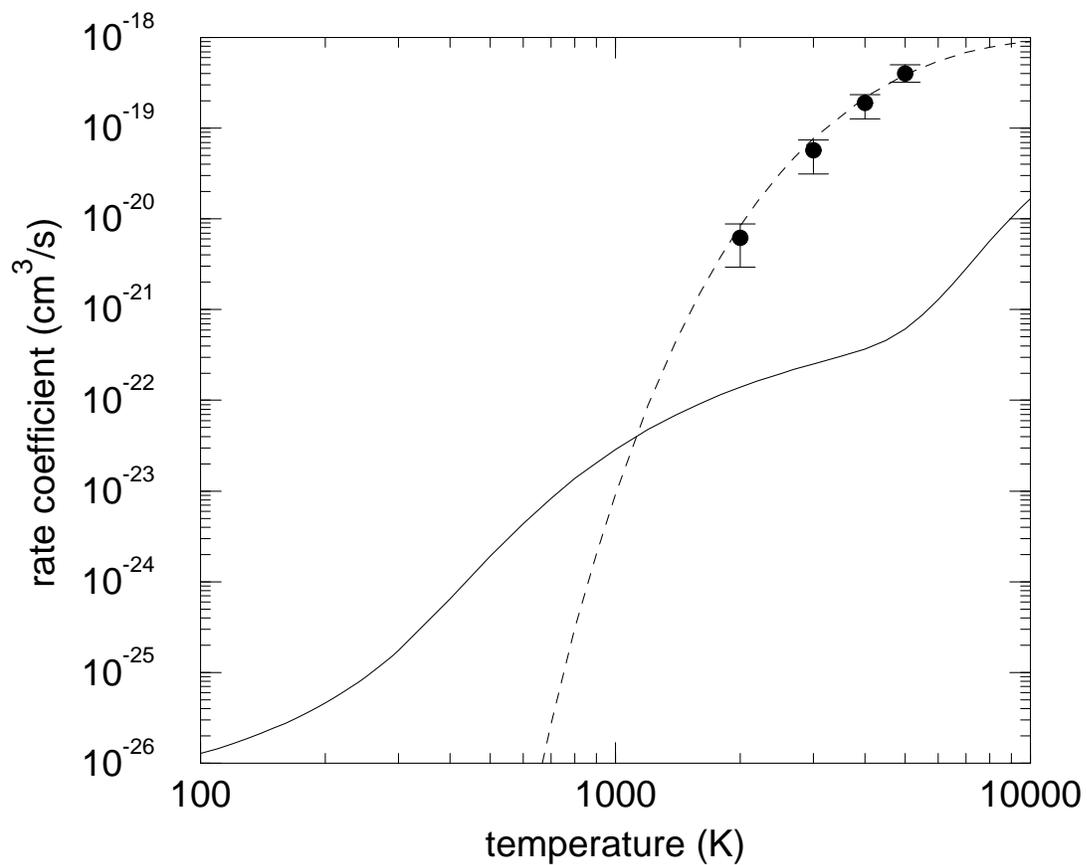}
\caption{\label{rates-figure}Comparison
of the calculated direct radiative association (solid line) and
inverse predissociation (dotted line) rate coefficients and the
experimental rate coefficient (points with error bars) of Myers and
Bartle.  }
\end{figure}

A rate coefficient
\begin{equation}
\label{invpre-rate-expt}
\alpha_{\rm Exp}(T)= 7.3\times10^{-18}
   \exp[(-14540 \pm 1100)/T] \; \mbox{cm}^3\mbox{s}^{-1}, \qquad
 280 \leq \lambda \leq  500~\mbox{nm},
\end{equation}
for the total radiative combination of
oxygen
\begin{equation}
\label{experiment-reaction}
\mbox{O}+\mbox{O} \rightarrow \mbox{O}_2 + h\nu
\end{equation}
between 2500~K and 3800~K was deduced by Myers and
Bartle~\cite{MyeBar68}.  In Table~\ref{rate-table} values of
$\alpha_{\rm Exp}(T)$ are given for various temperatures.  Our values
for the rate coefficient for inverse predissociation $\alpha_{\rm
IP}(T)$ are included in Table~\ref{rate-table}.  The rate coefficient
is fitted to within 10\% for $T\leq 5000$~K by the expression
\begin{equation}
\alpha_{\rm IP}(T) \sim  6.4\times10^{-18}
  \exp(-13\,455/T) \; \mbox{cm}^3\mbox{s}^{-1}, \quad
    T \leq 5000~\mbox{K},
\end{equation}
and to within 5\% for $ 5\,000 \leq T \leq 15\,000$~K by
\begin{equation}
\alpha_{\rm IP}(T) \sim (-5.6\times10^{-19}
        + 2.4\times10^{-22}T
          - 9.6\times10^{-27}T^2)\; \mbox{cm}^3\mbox{s}^{-1}, \quad
   5\,000 \leq T \leq 15\,000~\mbox{K} .
\end{equation}
In Fig.~\ref{rates-figure} the calculated direct $\alpha_{\rm D}$ and
inverse predissociation $\alpha_{\rm IP}$ rates are presented and
compared with the $\alpha_{\rm Exp}$.  The agreement is very good,
confirming that the formation of molecular oxygen from hot oxygen
atoms at low densities proceeds by inverse predissociation through the
$\mbox{B}\,{}^3\Sigma_u^-$ state following approach along the
$1\,{}^3\Pi_u$ state.

\section*{Acknowledgements}
We would like to thank our colleagues B. Zygelman, K. Yoshino,
J.-H. Yee, Y. Sun, R.~D. Sharma, R.~S.  Friedman, M.~L. Du, and A.~C.
Allison for sharing details of calculations and for helpful
discussions.  J.~Yeh, formerly of Harvard University, assisted in the
computation of the inverse predissociation rate coefficients and P.~C.
Stancil and Z.~Fan, formerly of the Harvard-Smithsonian Center for
Astrophysics, made improvements to the program used to calculate the
direct radiative association cross sections.  JFB would like to thank
P.~S.  Julienne and R.~E.  Meyerott for helpful discussions.  This
work has been supported by the National Science Foundation, Division
of Atmospheric Sciences under Grant No.  93-20175 and by grants for
the Institute for Theoretical Atomic and Molecular Physics to the
Smithsonian Astrophysical Observatory and Harvard University.


\clearpage
\begin{table}
\caption{\label{line-strength-table}Line strengths for transitions from
the $1\,{}^3\Pi_u$ to the $\mbox{X}\,{}^3\Sigma_g^-$ state.}
\begin{center}
\begin{tabular}{ccc}
  $J''$    &  Branch  &   $S_{J'',J}/(2S+1)$\\
\hline
  $J-1$    &  $R$     &   $J+1$  \\
  $J$      &  $Q$     &   $2J+1$ \\
  $J+1$    &  $P$     &   $J$    \\
\end{tabular}
\end{center}
\end{table}


\begin{table}
\caption{\label{rate-table}Rate coefficients
in $\mbox{cm}^3\mbox{s}^{-1}$, as a function of temperature $T$, in
$K$, for direct radiative association
process~(\protect\ref{direct-reaction}), inverse
predissociation~(\protect\ref{indirect-reaction}), and the
experimentally derived value for the total
process~(\protect\ref{experiment-reaction}). Numbers in square
brackets represent powers of 10.}
\begin{center}
\begin{tabular}{cccc}
\multicolumn{2}{c}{} &\multicolumn{1}{c}{Inverse} &
 \multicolumn{1}{c}{Experiment, Ref.~\protect\cite{MyeBar68}} \\
\multicolumn{1}{c}{$T(K)$} &
 \multicolumn{1}{c}{Direct} &
 \multicolumn{1}{c}{Predissociation} &
 \multicolumn{1}{c}{$\alpha_{\rm Exp}$, Eq.~(\protect\ref{invpre-rate-expt})}\\
\hline
15    &  6.0[-27] &          &          \\
30    &  6.4[-27] &          &          \\
50    &  7.4[-27] &          &          \\
100   &  1.3[-26] &          &          \\
200   &  4.6[-26] &          &          \\
300   &  1.7[-25] &          &          \\
400   &  6.5[-25] &          &          \\
500   &  1.9[-24] &          &          \\
600   &  4.4[-24] & 1.1[-27] &          \\
700   &  8.3[-24] & 2.7[-26] &          \\
800   &  1.4[-23] & 3.1[-25] &          \\
900   &  2.1[-23] & 2.0[-24] &          \\
1000  &  2.9[-23] & 9.2[-24] &          \\
1200  &  4.7[-23] & 8.9[-23] &          \\
1400  &  6.9[-23] & 4.5[-22] &          \\
1600  &  9.2[-23] & 1.5[-21] &          \\
1800  &  1.2[-22] & 3.9[-21] &          \\
2000  &  1.4[-22] & 8.4[-21] & 5.1[-21] \\
2200  &  1.6[-22] & 1.5[-20] & 9.8[-21] \\
2500  &  2.0[-22] & 3.2[-20] & 2.2[-20] \\
2700  &  2.2[-22] & 4.8[-20] & 3.3[-20] \\
3000  &  2.5[-22] & 7.8[-20] & 5.7[-20] \\
4000  &  3.7[-22] & 2.2[-19] & 1.9[-19] \\
5000  &  6.1[-22] & 3.9[-19] & 4.0[-19] \\
10000 &  1.7[-20] & 9.0[-19] &          \\
15000 &  7.3[-20] & 9.3[-19] &          \\
\end{tabular}
\end{center}
\end{table}
\clearpage
\begin{table}
\caption{\label{A-table}Transition probabilities
$A({v'},0)$ for emission from the $\mbox{B}\,{}^3\Sigma_u^-$ to
$\mbox{X}\,{}^3\Sigma_g^-$ states in units of $10^{7}\mbox{s}^{-1}$
for various values of the $\mbox{B}\,{}^3\Sigma_u^-$ vibrational
quantum number ${v'}$ with ${N'}=0$.}
\begin{center}
\begin{tabular}{ccccc}
\multicolumn{1}{c}{${v'}$} & \multicolumn{1}{c}{$A({v'},0)$} &
\multicolumn{1}{c}{ } &
\multicolumn{1}{c}{${v'}$} & \multicolumn{1}{c}{$A({v'},0)$} \\
\hline
0 & 1.816 & & 11 & 1.750 \\
1 & 1.916 & & 12 & 1.586 \\
2 & 2.006 & & 13 & 1.393 \\
3 & 2.078 & & 14 & 1.184 \\
4 & 2.130 & & 15 & 0.9709 \\
5 & 2.157 & & 16 & 0.7833 \\
6 & 2.157 & & 17 & 0.6162 \\
7 & 2.129 & & 18 & 0.4710 \\
8 & 2.073 & & 19 & 0.3414 \\
9 & 1.992 & & 20 & 0.2254 \\
10& 1.883 & & 21 & 0.1195
\end{tabular}
\end{center}
\end{table}

\clearpage
\clearpage
\clearpage

\end{document}